\documentclass[showpacs,amsmath,amssymb,11pt]{revtex4}

\allowdisplaybreaks
\usepackage{epsfig}

\begin{document}

\def\aprge{\buildrel > \over {_{\sim}}}
\def\aprle{\buildrel < \over {_{\sim}}}

\def\etal{{\it et.~al.}}
\def\ie{{\it i.e.}}
\def\eg{{\it e.g.}}

\def\bwt{\begin{widetext}}
\def\ewt{\end{widetext}}
\def\be{\begin{equation}}
\def\ee{\end{equation}}
\def\bea{\begin{eqnarray}}
\def\eea{\end{eqnarray}}
\def\bean{\begin{eqnarray*}}
\def\eean{\end{eqnarray*}}
\def\bary{\begin{array}}
\def\eary{\end{array}}
\def\bi{\bibitem}
\def\bit{\begin{itemize}}
\def\eit{\end{itemize}}

\def\lan{\langle}
\def\ran{\rangle}
\def\lra{\leftrightarrow}
\def\la{\leftarrow}
\def\ra{\rightarrow}
\def\dash{\mbox{-}}
\def\ol{\overline}

\def\ub{\ol{u}}
\def\db{\ol{d}}
\def\sb{\ol{s}}
\def\cb{\ol{c}}

\def\re{\rm Re}
\def\im{\rm Im}

\def \b{{\cal B}}
\def \ca{{\cal A}}
\def \ko{K^0}
\def \ok{\overline{K}^0}
\def \s{\sqrt{2}}
\def \st{\sqrt{3}}
\def \sx{\sqrt{6}}
\title{The Explicit Derivation of QED Trace Anomaly in Symmetry-Preserving Loop Regularization at One Loop Level}
\author{Jian-Wei Cui}
\affiliation{Center for High Energy Physics\\ Department of Engineering Physics\\ Tsinghua University, Beijing
100084, China.\\ jwcui@mail.tsinghua.edu.cn}
\author{Yong-Liang Ma}
\affiliation{Department of Physics, Nagoya University, Nagoya, 464-8602, Japan.\\
ylma@hken.phys.nagoya-u.ac.jp}
\author{Yue-Liang Wu}
\affiliation{ Kavli Institute for Theoretical Physics China (KITPC)
\\ Key Laboratory of Frontiers in Theoretical Physics \\ Institute of
Theoretical Physics, Chinese Academy of Science,   Beijing, 100190,
China \\ ylwu@itp.ac.cn}
\date{\today}
\begin{abstract}
The QED trace anomaly is calculated at one-loop level based on the loop regularization method which is realized in 4-dimensional spacetime and preserves gauge symmetry and Poincare symmetry in spite of the introduction of two mass scales, namely the ultraviolet (UV) cut-off $M_c$ and infrared (IR) cut-off $\mu_s$. It is shown that the dilation Ward identity which relates the three-point diagrams with the vacuum polarization diagrams gets the standard form of trace anomaly through quantum corrections in taking the consistent limit $M_c\to \infty$ and $\mu_s = 0$ which recovers the original integrals. This explicitly demonstrates that the loop regularization method is indeed a self-consistent regularization scheme which is applicable to the calculations not only for the chiral anomaly but also for the trace anomaly, at least at one-loop level. It is also seen that the consistency conditions which relates the tensor-type and scalar-type irreducible loop integrals (ILIs) are crucial for obtaining a consistent result. As a comparison, we also present the one-loop calculations by using the usual Pauli-Villars regularization and the dimensional regularization.
\end{abstract}




\maketitle

\section{introduction}

Symmetry and symmetry breaking play important roles in elementary particle physics. According to the Noether's theorem, if a system is invariant under a continue global transformation, there is a conserved current corresponding to this transformation. Although the Noether's theorem is an exact one, it comes into exist at the classical level. Sometimes the consequences of the Noether's theorem are violated by quantum corrections and are related with anomaly. A well known example of anomaly is the chiral anomaly which is due to Adler, Bell and Jackiw~\cite{Bell:1969ts,Adler:1969gk}. In the evaluation of the anomaly, an unavoidable procedure is to regularize the loop integrals, therefore a gauge symmetry-preserving regularization is crucial for the calculation of quantum correction induced anomaly. Meanwhile, a calculation of quantum correction induced anomaly provides a useful laboratory for the check of the consistence of a new regularization prescription.

Recently, a new regularization scheme named loop regularization was proposed\cite{Wu:2002xa,Wu:2003dd}. It has been explicitly proved at one-loop level that the loop regularization can preserve the non-ableian gauge symmetry~\cite{Cui:2008uv} as well as supersymmetry~\cite{Cui:2008bk}. Adopting such a symmetry-preserving loop regularization, we have studied the chiral anomaly systematically~\cite{Ma:2005md}. It was found that the two scales in the loop regularization, UV scale $M_c$ and IR scale $\mu_s$, play important roles in understanding the chiral anomaly. Under the condition $M_c\rightarrow\infty$ which is applicable for the renormalizability of QED and a symmetrical treatment of all the three currents, the chiral anomaly in the massless QED can consistently be obtained in case of $\mu_s = 0$, while the massless QED could be free from anomaly if keeping $\mu_s \neq 0$. For the massive QED, the chiral anomaly can be obtained with $\mu_s = 0$, while if taking $\mu_s \gg m$, the chiral anomaly would also be absent. In Ref.~\cite{Ma:2006yc}, based on the loop regularization method, the radiatively induced Lorentz and CPT violating Chern-Simons term in QED was calculated and a consistent result is obtained when simultaneously combining the evaluation for the chiral anomaly with ensuring the vector current conserved. In Refs.~\cite{Tang:2008ah,Tang:2010cr}, the loop regularization method has been applied to study the gauge theory coupled to gravitation.

In this paper, we will focus on another well-known anomaly, namely trace anomaly, which causes the violation of the dilation Ward identity due to the scaling invariance. The trace anomaly problem was firstly studied in Ref.~\cite{Callan:1970yg} in the framework of scalar field theory, and it was subsequently analyzed in various aspects~\cite{Coleman:1970je,Chanowitz:1972da,Adler:1976zt}. In this paper, we shall calculate the trace anomaly in the framework of QED by using the symmetry-preserving loop regularization~\cite{Wu:2002xa,Wu:2003dd}. Since the standard form of the trace anomaly in QED is known, for example Ref.~\cite{Chanowitz:1972da}, our study can be regarded as a further independent check of the consistency of the loop regularization prescription. This comes to our main purpose in the present paper. We shall show that with the symmetry-preserving loop regularization one can obtain the standard form of the trace anomaly when taking $M_c \rightarrow \infty$ and $\mu_s = 0$, which is required for recovering the original Feynman loop integrals and also applicable to QED as it is a renormalizable and IR divergence free theory. It is also seen that the consistency conditions relating the tensor type and the scalar type irreducible loop integrals, which have been verified in the loop regularization method, are crucial for the anomaly evaluation. From our explicit calculations, we arrive at the conclusion again that the loop regularization is indeed a consistent symmetry-preserving regularization.

The paper is organized as follows: In Sec.~\ref{sec:ward}, we discuss the dilation ward identity. Sec.~\ref{sec:loop} presents our detailed calculations for the trace anomaly based on the loop regularization method. In Sec.~\ref{sec:PV and Di}, we give, for a comparison, the calculations by using the Pauli-Villars and dimensional regulations. The last section is our conclusions.

\section{The dilation Ward identity}

\label{sec:ward}

The dilation transformation is a rescaling of the space-time coordinate
\begin{eqnarray}
x^\mu \rightarrow x^{\prime\mu} = \lambda x^\mu,
\end{eqnarray}
with $\lambda$ is a real constant. Under the dilation transformation, a field $\Phi(x)$ transforms as
\begin{eqnarray}
\delta_D\Phi(x) & = & i[\mathbf{D},\Phi(x)] =
(x\cdot\partial+d)\Phi(x),
\end{eqnarray}
where $d$ is the scale dimension of the field $\Phi(x)$ and classically, it is the mass dimension of the field $\Phi(x)$, i.e., one for scalar field, three half for fermion field and others for composite operators. $\mathbf{D}$ is the generator of the dilation transformation.

It was proved that~\cite{Callan:1970ze} the current $D_\mu$ generated by the dilation transformation is related to the improved symmetric energy-momentum tensor $\theta_{\mu\nu}$ by
\begin{eqnarray}
D_{\mu}(x) & = & x^\nu \theta_{\mu\nu},
\end{eqnarray}
then the trace of $\theta_{\mu\nu}$ is the divergence of the dilation current
\begin{eqnarray}
\theta^{\mu}_{\mu}(x) & = &
\partial_{\mu}D^{\mu}(x).\label{divCurrent}
\end{eqnarray}
By using the canonical commutative relations, one can show that the dilation charge $D(x_0)=\int{d^3x}D^0(x_i,x_0)$ can be taken as the generator of dilation transformation, that is,
\begin{eqnarray}
[D(x_0),\Phi(\vec{x},x_0)] =
-i(x\cdot\partial+d)\Phi(x).\label{charge-generator}
\end{eqnarray}
With respect to this commutation relation, we can derive a simple dilation Ward identity (see Appendix~\ref{app:a})
\begin{eqnarray}
(2-p\cdot\frac{\partial}{\partial{p}})\Pi_{\mu\nu}(p,-p) & = &
\Delta_{\mu\nu}(p,-p), \label{classical-Ward}
\end{eqnarray}
where
\begin{eqnarray}
\Pi_{\mu\nu}(p,-p) & = &
i\int{d^4x}e^{ip\cdot{x}}\langle{T^\ast}(J_{\mu}(x)J_{\nu}(0))\rangle,
\label{vacuumpolari} \\
\Delta_{\mu\nu}(p,-p) & = &
\int{d^4x}{d^4y}e^{ip\cdot{y}}\langle{T^\ast}(\theta^{\lambda}_{\lambda}(x)J_{\mu}(y)J_{\nu}(0))\rangle,
\label{vertex}
\end{eqnarray}
which can be diagrammatically illustrated in Fig.~\ref{fig:2and3point}.

\section{trace anomaly calculation with loop regularization}

\label{sec:loop}

The Ward identity (\ref{classical-Ward}) is a classical one and it is violated by quantum corrections~\cite{Chanowitz:1972da}. In this section, we will investigate this violation using symmetry-preserving loop regularization~\cite{Wu:2002xa,Wu:2003dd}. Our discussion is specified in the framework of QED with lagrangian
\begin{eqnarray}
{\cal L} & = & \bar{\psi}\gamma^\mu(i\partial_\mu-e{A}_\mu)\psi-m\bar{\psi}\psi-\frac{1}{4}F^{\mu\nu}F_{\mu\nu}.
\label{QED}
\end{eqnarray}
From this lagrangian, the symmetric energy-momentum tensor can be easily got as~\cite{Adler:1976zt}
\begin{eqnarray}
\theta_{\mu\nu} & = & {i\over4}\Big[\bar{\psi}\gamma_\mu(\overrightarrow{\partial}_\nu+ieA_\nu)\psi+\bar{\psi}\gamma_\nu(\overrightarrow{\partial}_\mu+ieA_\mu)\psi-\bar{\psi}\gamma_\mu(\overleftarrow{\partial}_\nu-ieA_\nu)\psi-\bar{\psi}\gamma_\nu(\overleftarrow{\partial}_\mu-ieA_\mu)\psi \Big]\nonumber\\
& &
+{1\over4}g_{\mu\nu}F^{\lambda\sigma}F_{\lambda\sigma}-F^{\lambda}_{\mu}F_{\lambda\nu}.
\end{eqnarray}
Then, by using the equation of motion, we get
\begin{eqnarray}
\theta_{\mu}^\mu & = & g_{\mu\nu}\theta^{\mu\nu} = m\bar{\psi}\psi.\label{Eq. cla tra}
\end{eqnarray}

\begin{figure}
\begin{center}
\includegraphics[scale=0.6]{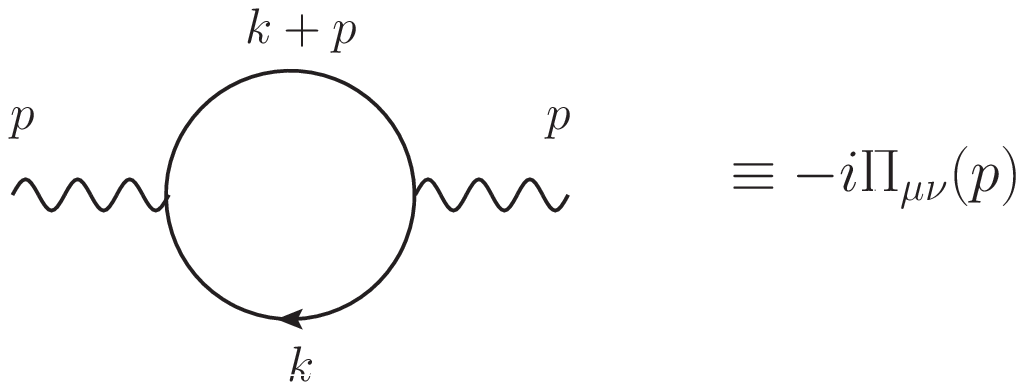} \hspace{1.2cm} \includegraphics[scale=0.6]{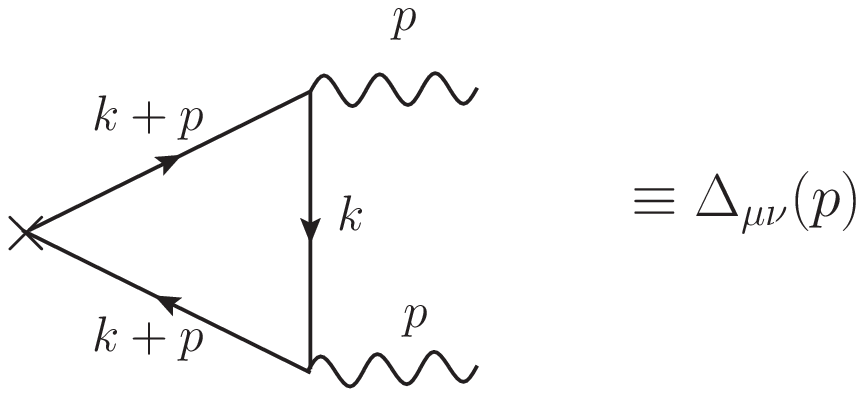}
\end{center}
\caption{One loop diagrammatical representation  of $\Pi_{\mu\nu}$ and $\Delta_{\mu\nu}$.} \label{fig:2and3point}
\end{figure}

For studying the quantum corrections to the dilation Ward identity (\ref{classical-Ward}), one should consider the diagrams depicted in Fig.~\ref{fig:2and3point}. For the vacuum polarization diagram, after some algebra, one has
\begin{eqnarray}
-i\Pi_{\mu\nu} & = &
(-)\int\frac{d^4k}{(2\pi)^4}{\rm
Tr}[e\gamma_{\mu}\frac{i}{k\hspace{-0.2cm}\slash-m}e\gamma_{\nu}\frac{i}{(k\hspace{-0.2cm}\slash+p\hspace{-0.2cm}\slash)-m}] \nonumber\\
& = &
4e^2\int_0^1dx\Big[2I_{2\mu\nu}(m)-I_2(m)g_{\mu\nu}+2x(1-x)(p^2g_{\mu\nu}-p_{\mu}p_{\nu})I_0(m)\Big],
\end{eqnarray}
where
\begin{eqnarray}
I_{2\mu\nu}(m) & = & \int\frac{d^4k}{(2\pi)^4}\frac{k_\mu k_\nu}{(k^2-M^2)^2}, \nonumber\\
I_2(m) & = & \int\frac{d^4k}{(2\pi)^4}\frac{1}{k^2-M^2}, \nonumber\\
I_0(m) & = & \int\frac{d^4k}{(2\pi)^4}\frac{1}{(k^2-M^2)^2},
\end{eqnarray}
with $M^2 = m^2 - x(1-x)p^2$. The number subscript of $I$ stands for the degree of divergence of the integral and specifically, "$2$" means the corresponding integrals are quadratically divergent, and "$0$" means the corresponding integrals are logarithmically divergent. After applying the loop regularization\cite{Wu:2002xa}, we yield the consistency condition
\begin{eqnarray}
2I_{2\mu\nu}^R(m) & = & g_{\mu\nu}I_2^R(m), \label{LR1}
\end{eqnarray}
and the regularized vacuum polarization two-point function
\begin{eqnarray}
-i\Pi_{\mu\nu}^R & = &
8e^2(p^2g_{\mu\nu}-p_{\mu}p_{\nu})\int_0^1dxx(1-x)I_0^R(m).\label{LRP}
\end{eqnarray}
The explicit forms of $I_2^R(m)$ and $I_0^R(m)$ were given as ~\cite{Wu:2002xa}
\begin{eqnarray}
I_2^R & = &
-\frac{i}{16\pi^2}\Big\{M_c^2-\mu^2\Big[\ln\frac{M_c^2}{\mu^2}-\gamma_\omega+1+
y_2(\frac{\mu^2}{M_c^2})\Big]\Big\},\label{I2R}\\
I_0^R & = &
\frac{i}{16\pi^2}\Big[\ln\frac{M_c^2}{\mu^2}-\gamma_\omega
+y_0(\frac{\mu^2}{M_c^2})\Big],\label{I0R}
\end{eqnarray}
with $ \mu^2 = \mu_s^2 + M^2$, $\gamma_w = \gamma_E = 0.5772\cdots$, and
\begin{eqnarray}
& & y_0 (x) = \int_0^x d \sigma\ \frac{1 - e^{-\sigma}
}{\sigma},\nonumber\\
& & y_1 (x)  = \frac{e^{-x} - 1 + x}{x}, \nonumber\\
& & y_2(x) = y_0(x) - y_1(x),
\end{eqnarray}
where $\mu_s $ and $M_c$ are introduced in the loop regularization~\cite{Wu:2002xa,Wu:2003dd} and, roughly speaking, they play the roles of UV and IR cutoff, respectively. Because the QED is a renormalizable and IR divergent free theory, we can safely set $M_c$ to be infinity and $\mu_s$ to be zero, i.e., $M_c\rightarrow\infty$ and $\mu_s = 0$.

With the above analysis, after a simply algebra, the left hand side (l.h.s.) of $(\ref{classical-Ward})$ can be simply written as
\begin{eqnarray}
(2-p\cdot\frac{\partial}{\partial{p}})\Pi_{\mu\nu}^R(p,-p) & = &
32ie^2(p^2g_{\mu\nu}-p_{\mu}p_{\nu})p^2\int_0^1dxx^2(1-x)^2I_{-2}^R(m). \label{LRPion}
\end{eqnarray}
To get the above relation, we have used the following expressions
\begin{eqnarray}
& & I_{-2}^R = -\frac{i}{16\pi^2}\frac{1}{2\mu^2}[1-y_{-2}(\frac{\mu^2}{M_c^2})]\label{I-2R}, \nonumber\\
& & y_{-2} (x) = 1 - e^{-x}, \nonumber
\end{eqnarray}
and relation
\begin{eqnarray}
\frac{\partial}{\partial{p_{\mu}}}I_0^R(m) =
-4x(1-x)p^{\mu}I_{-2}^R(m)\ .
\end{eqnarray}

Now let us turn to the three-point function (\ref{vertex}) which is illustrated in Fig.~\ref{fig:2and3point}. Including the crossing diagram, we get the expression
\begin{eqnarray}
\Delta_{\mu\nu}(p,-p) & = & (-)2\int\frac{d^4k}{(2\pi)^4}{\rm
Tr}\Big[e\gamma_{\mu}\frac{i}{k\hspace{-0.2cm}\slash-m}e\gamma_{\nu}\frac{i}{(k\hspace{-0.2cm}\slash+p\hspace{-0.2cm}\slash)-m}m\frac{i}{(k\hspace{-0.2cm}\slash+p\hspace{-0.2cm}\slash)-m}\Big]\nonumber\\
& = & ie^2\int_0^1dx\Big\{16xm^2\Big[4I_{0\mu\nu}(m)-g_{\mu\nu}I_0(m)\Big]\nonumber\\
& &
\;\;\;\;\;\;\;\;\;\;\;\;\;\;\;\;\;\; - 64x^2(1-x)m^2(p_{\mu}p_{\nu}-g_{\mu\nu}p^2)I_{-2}(m)\nonumber\\
& & \;\;\;\;\;\;\;\;\;\;\;\;\;\;\;\;\;\; +
16x(1-x)(1-2x)m^2g_{\mu\nu}p^2I_{-2}(m)\Big\}.
\end{eqnarray}
After applying loop regularization with the consistency condition
\begin{eqnarray}
I_{0\mu\nu}^R(m)=\frac{1}{4}g_{\mu\nu}I_0^R(m), \label{LR2}
\end{eqnarray}
we arrive at
\begin{eqnarray}
\Delta_{\mu\nu}^R(p,-p) & = & -64ie^2(p_{\mu}p_{\nu}-g_{\mu\nu}p^2)\int_0^1dxx^2(1-x)m^2I_{-2}^R(m)\nonumber\\
& & + 16ie^2m^2g_{\mu\nu}p^2\int_0^1dxx(1-x)(1-2x)I_{-2}^R(m).
\end{eqnarray}
From this expression, at the first sight, one may think that, for the triangle graph evaluated by loop regularization, the gauge invariance is not preserved due to the second term. In fact, this is not the case if one realizes the following general relation for the Feynman parameter integral
\begin{eqnarray}
\int_0^1dx(1-2x)f[x(1-x)] & = & 0,
\end{eqnarray}
which can be proved by changing the integration variable from $x$ to $1-x$, where $(1-2x)= (1-x) -x $ is an odd function under such a change. We then arrive at a gauge invariant result
\begin{eqnarray}
\Delta_{\mu\nu}^R(p,-p) & = &
-32ie^2m^2(p_{\mu}p_{\nu}-g_{\mu\nu}p^2)\int_0^1dxx(1-x)I_{-2}^R(m),\label{LRDelt}
\end{eqnarray}
which is similar to Eq.~(\ref{LRPion}).

Takeing the limit $M_c\rightarrow\infty$ and using the explicit form of $I_{-2}^R$, we have
\begin{eqnarray}
(2-p\cdot\frac{\partial}{\partial{p}})\Pi_{\mu\nu}^R(p,-p)
& = & -\frac{e^2}{\pi^2}(p_{\mu}p_{\nu}-g_{\mu\nu}p^2)\int_0^1dxx^2(1-x)^2\frac{p^2}{\mu_s^2+M^2}, \nonumber\\
\Delta_{\mu\nu}^R(p,-p) & = &
-\frac{e^2}{\pi^2}(p_{\mu}p_{\nu}-g_{\mu\nu}p^2)\int_0^1dxx(1-x)\frac{m^2}{\mu_s^2+M^2}.
\end{eqnarray}
From these two equations, we obtain their difference
\begin{eqnarray}
\Delta_{\mu\nu}^R(p,-p)-(2-p\cdot\frac{\partial}{\partial{p}})\Pi_{\mu\nu}^R(p,-p)
&  = &
-\frac{e^2}{\pi^2}(p_{\mu}p_{\nu}-g_{\mu\nu}p^2)\int_0^1dxx(1-x)\frac{m^2-x(1-x)p^2}{\mu_s^2+m^2-x(1-x)p^2}.\nonumber
\end{eqnarray}
As QED is an IR divergence free theory, we safely take $\mu_s=0$, which leads to the following simple relation
\begin{eqnarray}
\Delta_{\mu\nu}^R(p,-p)-(2-p\cdot\frac{\partial}{\partial{p}})\Pi_{\mu\nu}^R(p,-p)
& = & -\frac{e^2}{6\pi^2}(p_{\mu}p_{\nu}-g_{\mu\nu}p^2)\nonumber\\
& = &
-(2-p\cdot\frac{\partial}{\partial{p}})\Pi_{\mu\nu}^R(p,-p,m=0).
\end{eqnarray}
Thus the dilation Ward identity at one-loop order is given by
\begin{eqnarray}
\Delta_{\mu\nu}^R(p,-p) & = & (2-p\cdot\frac{\partial}{\partial{p}})\Pi_{\mu\nu}^R(p,-p)-(2-p\cdot\frac{\partial}{\partial{p}})\Pi_{\mu\nu}^R(p,-p,m=0) \nonumber\\
& = &
(2-p\cdot\frac{\partial}{\partial{p}})\Pi_{\mu\nu}^R(p,-p)-\frac{e^2}{6\pi^2}(p_{\mu}p_{\nu}-g_{\mu\nu}p^2),\label{anoma-1}
\end{eqnarray}
which demonstrates that the dilation Ward identity (\ref{classical-Ward}) is violated by quantum corrections and the anomaly arises from the quantum effects. One can also see that our conclusion is exactly the same as that given in Ref.~\cite{Chanowitz:1972da}.

It is interesting to note that the anomalous term in Eq.~(\ref{anoma-1}) actually arises from the photon polarization $\Pi_{\mu\nu}$. This can be seen explicitly in the case of the massless QED. In the massless case, the classical ward identity becomes
\begin{eqnarray}
\Delta_{\mu\nu}(p,-p)=(2-p\cdot\frac{\partial}{\partial{p}})\Pi_{\mu\nu}(p,-p,m=0) = 0.
\end{eqnarray}
While one-loop contribution to the vacuum polarization gives
\begin{eqnarray}
(2-p\cdot\frac{\partial}{\partial{p}})\Pi_{\mu\nu}^R(p,-p,m=0)=
32ie^2(p^2g_{\mu\nu}-p_{\mu}p_{\nu})p^2\int_0^1dxx^2(1-x)^2I_{-2}^R(0),
\end{eqnarray}
which leads, in the limits $M_c\rightarrow\infty$ and $\mu_s=0$, to the following result
\begin{eqnarray}
(2-p\cdot\frac{\partial}{\partial{p}})\Pi_{\mu\nu}^R(p,-p,m=0) & = &
32ie^2(p^2g_{\mu\nu}-p_{\mu}p_{\nu})p^2\int_0^1dxx^2(1-x)^2\Big\{-\frac{i}{16\pi^2}\frac{1}{2[-p^2x(1-x)]}\Big\}\nonumber\\
& = & \frac{e^2}{6\pi^2}(p_{\mu}p_{\nu}-p^2g_{\mu\nu}).,
\end{eqnarray}
It is exactly the anomaly in the l.h.s. of (\ref{anoma-1}).

In an operator form, the anomalous Ward identity (\ref{anoma-1}) can be written as
\begin{eqnarray}
\theta^\mu_\mu & = &
m\bar{\psi}\psi+\frac{2\alpha_{\rm em}}{3\pi}(\frac{1}{4}F_{\mu\nu}F^{\mu\nu}),
\end{eqnarray}
with $\alpha_{\rm em}$ as the fine-structure constant of QED. This one-loop level result agrees with the all order result given in Ref.~\cite{Adler:1976zt}. To all orders in perturbation theory, it has been proved that the trace anomaly can be expressed in the following form
\begin{eqnarray}
\theta^\mu_\mu & = & \left( 1+\delta(\alpha_{\rm em})\right) m_0\bar{\psi}\psi+\frac{\beta(\alpha_{\rm em})}{\alpha_{\rm em}}N(\frac{1}{4}F_{\mu\nu}F^{\mu\nu}),\label{all order}
\end{eqnarray}
where $N(\frac{1}{4}F_{\mu\nu}F^{\mu\nu})$ is a subtracted form of the operator $\frac{1}{4}Z_3^{-1}F_{\mu\nu}F^{\mu\nu}$ which satisfies Eq.~(2.3) in Ref.~\cite{Adler:1976zt} after composed operator renormalization, and $Z_3$ is the wave-function renormalization constant of photon. $\delta(\alpha_{\rm em}) \equiv \frac{m}{m_0}\frac{\partial m_0}{\partial m}-1$ is the QED anomalous mass dimension function, and $\beta(\alpha_{\rm em}) \equiv m\frac{\partial \alpha_{\rm em}}{\partial m}$ is the QED $\beta$ function in the on-shell scheme, with $m_0$ and $m$ as the bare and physical masses of electron, respectively. Here we would like to mention that generally in the one coupling constant theory the $\beta$ function starts to depend on the renormalization scheme from the three-loop level, and at present the QED $\beta$ function in the on-shell scheme is obtained at four-loop order in Ref.~\cite{Broadhurst:1992za}
. Eq.~(\ref{all order}) means that the trace anomaly is proportional to $\beta$-function to all orders. From the above one-loop order result of trace anomaly, we can obtain the well-known lowest order QED $\beta$-function~\cite{Itzykson:1980rh} as $\beta(\alpha_{\rm em})=\frac{2\alpha^2_{\rm em}}{3\pi}$.


\section{Trace anomaly with other regularization schemes at one-loop level}

\label{sec:PV and Di}


In Pauli-Villars regularization, the regularized lagrangian of QED is
\begin{eqnarray}
\mathcal{L}^R=\sum_{i=0}^{n}C_i\{\bar{\psi_i}\gamma^\mu(i\partial_\mu-eA_\mu)\psi_i-m_i\bar{\psi_i}\psi\}.
\end{eqnarray}
It is enough to choose $n = 2$ to make the whole theory finite, and $C_i$ and $m_i$ can be specialized as
\begin{eqnarray}
& & C_0 = 1,\;\;\; C_1 = 1, \;\;\; C_2 = -2,\nonumber\\
& & m_0^2 = m^2, \;\;\; m_1^2 = m^2 + 2\Lambda^2, \;\;\; m_2 = m^2 +
\Lambda^2, \label{Pauli-condition}
\end{eqnarray}
with $\Lambda$ as a constant representing the UV cutoff which should be set to infinity in the final result. Then the regularized
$\Delta_{\mu\nu}^R$ is
\begin{eqnarray}
\Delta_{\mu\nu}^R(p,-p) =
-2\sum_{i=0}^{2}C_i\int\frac{d^4k}{(2\pi)^4}{\rm
Tr}[e\gamma_{\mu}\frac{i}{k\hspace{-0.2cm}\slash-m_i}e\gamma_{\nu}\frac{i}{(k\hspace{-0.2cm}\slash+p\hspace{-0.2cm}\slash)-m_i}m_i\frac{i}{(k\hspace{-0.2cm}\slash+p\hspace{-0.2cm}\slash)-m_i}].
\end{eqnarray}
After Feynman parametrization, a direct one-loop calculation gives
\begin{eqnarray}
\Delta_{\mu\nu}^R(p,-p) & =
 & \frac{e^2}{4\pi^2}g_{\mu\nu}\sum_{i=0}^{2}C_im_i^2-\frac{e^2}{4\pi^2}(p_{\mu}p_{\nu}-g_{\mu\nu}p^2)\sum_{i=0}^{2}C_im_i^2
\int_{0}^{1}dx\frac{x(1-x)}{m_i^2-x(1-x)p^2}.\label{DeltPV}
\end{eqnarray}
By using the conditions (\ref{Pauli-condition}), the first term vanishes. It should be mentioned that in Ref.~\cite{Chanowitz:1972da} the mass terms is discarded directly without considering these condition because the coefficient $C_i$ is omitted there. Then after the Feynman parameter integral, we obtain
\begin{eqnarray}
\Delta_{\mu\nu}^R(p,-p) & = &
-\frac{e^2}{\pi^2}(p_{\mu}p_{\nu}-g_{\mu\nu}p^2)\sum_{i=0}^{2}C_i\frac{m_i^2}{p^2}(m_i^2A_i-1),\label{PVDelt}
\end{eqnarray}
where $A_i$ is given by
\begin{eqnarray}
A_i(p^2) & = &
\frac{2}{(p^4-4m_i^2p^2)^{1/2}}\ln\frac{p^2-(p^4-4m_i^2p^2)^{1/2}}{p^2+(p^4-4m_i^2p^2)^{1/2}}.
\end{eqnarray}
As required by the Pauli-Villar regularization, to get the meaningful result of the original theory we should take limit $\Lambda\rightarrow\infty$ in the expression (\ref{PVDelt}). So that in this limit case, all the heavy regulators give vanishing contributions, and the result can be written as
\begin{eqnarray}
\Delta_{\mu\nu}^R(p,-p) =
-\frac{e^2}{\pi^2}(p_{\mu}p_{\nu}-g_{\mu\nu}p^2)\frac{m^2}{p^2}(m^2A_0-1).
\end{eqnarray}

For the one-loop vacuum polarization diagram $\Pi_{\mu\nu}(p,-p)$, it has been calculated in Ref.~\cite{BD}. The result can be read
\begin{eqnarray}
\Pi_{\mu\nu}^R(p,-p) =
\frac{e^2}{12\pi^2}(p_{\mu}p_{\nu}-g_{\mu\nu}p^2)\Big\{\ln\frac{\Lambda^2}{m^2}-6\int_0^1dxx(1-x)\ln[1-\frac{p^2}{m^2}x(1-x)]\Big\},
\end{eqnarray}
from which we can easily obtain
\begin{eqnarray}
(2-p\cdot\frac{\partial}{\partial{p}})\Pi_{\mu\nu}^R(p,-p)=-\frac{e^2}{\pi^2}(p_{\mu}p_{\nu}-g_{\mu\nu}p^2)\Big[\frac{m^2}{p^2}(m^2A_0-1)-\frac{1}{6}\Big].\label{PVVP}
\end{eqnarray}
Combing (\ref{PVDelt}) and (\ref{PVVP}), we finally get
\begin{eqnarray}
\Delta_{\mu\nu}^R(p,-p)=(2-p\cdot\frac{\partial}{\partial{p}})\Pi^R_{\mu\nu}(p,-p)-\frac{e^2}{6\pi^2}(p_{\mu}p_{\nu}-p^2g_{\mu\nu}).
\end{eqnarray}
This means that the classical Ward identity (\ref{classical-Ward}) is violated by an anomalous term. Compared with Eq.~(\ref{anoma-1}), we find that the anomalous term calculated with Pauli-Villars regularization is the same as that obtained with loop regularization. In the massless QED case, we can get the same conclusion as in loop regularization.

For the calculation with dimensional regularization\cite{DR}, one can draw the same conclusion. In the calculation with loop regularization, it has been seen that the two consistency conditions (\ref{LR1}) and (\ref{LR2}) are crucial to get the consistent conclusion. In dimensional regularization, these two conditions are satisfied, therefore the conclusions of the forms (\ref{LRP}) and (\ref{LRDelt}) can directly be reached. The difference is that, the explicit forms of $I_0^R(m)$ and $I_{-2}^R(m)$ in dimensional regularization can be resulted from that in loop regularization by taking the limits $M_c\rightarrow\infty$ and $\mu_s=0$. We give the one-loop results with dimensional regularization here
\begin{eqnarray}
\Pi_{\mu\nu}^R(p,-p)&=&\frac{e^2}{12\pi^2}(p_{\mu}p_{\nu}-p^2g_{\mu\nu})\frac{1}{\epsilon}-\frac{e^2}{2\pi^2}(p_{\mu}p_{\nu}-p^2g_{\mu\nu})\int_0^1~dx~x(1-x){\rm ln}\frac{m^2-x(1-x)p^2}{\mu^2}\nonumber\\
\Delta_{\mu\nu}^R(p,-p)&=&-\frac{2e^2}{\pi^2}(p_{\mu}p_{\nu}-p^2g_{\mu\nu})\int_0^1~dx~x^2(1-x)\frac{m^2}{m^2-x(1-x)p^2}.
\end{eqnarray}
From these equations, the calculation of anomaly is straightforward. We notice here that the anomalous term only depends on the finite parts of the above expressions, so that no matter what subtraction scheme be used we can always obtain the following result
\begin{eqnarray}
& &(2-p\cdot\frac{\partial}{\partial{p}})\Pi^R_{\mu\nu}(p,-p)-\Delta_{\mu\nu}^R(p,-p)\nonumber\\
&=&\frac{e^2}{\pi^2}(p_{\mu}p_{\nu}-p^2g_{\mu\nu})\int_0^1~dx~\frac{-x^2(1-x)^2p^2+2x^2(1-x)m^2}{m^2-x(1-x)p^2}\nonumber\\
&=&\frac{e^2}{6\pi^2}(p_{\mu}p_{\nu}-p^2g_{\mu\nu}).
\end{eqnarray}

This subtraction scheme independent anomaly is the same as the one-loop results calculated with loop regularization and with Pauli-Villars regularization, and it demonstrates the self-consistency of applications of different methods.

\section{conclusions}

\label{sec:con}

In this paper, we have calculated the trace anomaly with loop regularization at one-loop level. The explicit demonstration has shown that the trace anomaly can be consistently obtained under the conditions $M_c\rightarrow\infty$ and $\mu_s = 0$ which are necessary for recovering original theory and are also applicable for our present calculations as QED is a renormalizable and IR divergence free theory. It has also been seen that the consistency conditions which are the direct deductions of non-abelian gauge symmetry are crucial for the anomaly evaluation. As the loop regularization preserves these conditions~\cite{Wu:2002xa,Wu:2003dd}, it is then applicable for the studies on both symmetries and anomalies of gauge theories.

We would like to mention that in low orders the anomalous dilation Ward identity (\ref{anoma-1}) can be rewritten in the form of the classical one with the dimension $d$ substituted by $d^\prime$ which is a sum of the classical dimension and the anomalous dimension~\cite{Coleman:1970je}. In the language of renormalization, this can be understood as follows: As the vacuum polarization can be incorporated into the normalization constant of photon wave function $Z_3$, the renormalized photon wave function has the same form as the bare one except a scale dependent normalization constant $Z_3$. Any regularization scheme will introduce at least one dimensional parameter into the bare theory, such as $M_c$ in loop regularization, $\Lambda$ in Pauli-Villar regularization and $\mu$ in dimensional regularization. It is exactly this dimensional parameter which breaks the tree-level dilation Ward identity. On the other side, it is well-known that an exact scale invariant theory should be massless, so that the scale violation of a massless theory may be understood as the emergence of mass. In this sense, the radiative corrections can be considered as a possible way to generate mass~\cite{Coleman:1973jx}.


\acknowledgments

\label{ACK}

This work of Y.W is supported in part by the National Nature Science Foundation of China (NSFC) under Grants No. 10975170, No. 10821504 and No. 10905084; and the Project of Knowledge Innovation Program (PKIP) of the Chinese Academy of Science. The work of Y.M is supported in part by Grant-in-Aid for Scientific Research on Innovative Areas (No. 2104) ``Quest on New Hadrons with Variety of Flavors'' from MEXT and the National Science Foundation of China (NNSFC) under grant No. 10905060. The work of J.C is supported in part by the China Postdoctoral Science Foundation (CPSF) under Grant No. 20090460364.

\appendix

\section{Derivation of Classical Ward Identity Eq.~(\ref{classical-Ward}).}

\label{app:a}

In this appendix, we shall derive the classical Ward identity eq.(\ref{classical-Ward}). Let us begin with eq.(\ref{vertex}) by considering the property of time order product, have
\begin{eqnarray}
\Delta_{\mu\nu}(p,-p)&=&\int{d^4x}{d^4y}e^{ip\cdot{y}}\langle{T^*}(\theta^{\lambda}_{\lambda}(x)J_{\mu}(y)J_{\nu}(0))\rangle\nonumber\\
&=&\int{d^4x}{d^4y}e^{ip\cdot{y}}\theta(x_0)\theta(x_0-y_0)\theta(y_0)\langle\theta^{\lambda}_{\lambda}(x)J_{\mu}(y)J_{\nu}(0)\rangle\nonumber\\
&&+\int{d^4x}{d^4y}e^{ip\cdot{y}}\theta(x_0)\theta(y_0-x_0)\langle J_{\mu}(y)\theta^{\lambda}_{\lambda}(x)J_{\nu}(0)\rangle\nonumber\\
&&+\int{d^4x}{d^4y}e^{ip\cdot{y}}\theta(-x_0)\theta(y_0)\langle J_{\mu}(y)J_{\nu}(0)\theta^{\lambda}_{\lambda}(x)\rangle\nonumber\\
&&+\int{d^4x}{d^4y}e^{ip\cdot{y}}\theta(-x_0)\theta(y_0-x_0)\theta(-y_0)\langle J_{\nu}(0)J_{\mu}(y)\theta^{\lambda}_{\lambda}(x)\rangle\nonumber\\
&&+\int{d^4x}{d^4y}e^{ip\cdot{y}}\theta(-x_0)\theta(x_0-y_0)\langle J_{\nu}(0)\theta^{\lambda}_{\lambda}(x)J_{\mu}(y)\rangle\nonumber\\
&&+\int{d^4x}{d^4y}e^{ip\cdot{y}}\theta(x_0)\theta(-y_0)\langle
\theta^{\lambda}_{\lambda}(x)J_{\nu}(0)J_{\mu}(y)\rangle.
\end{eqnarray}
with respect to the relation eq.(\ref{divCurrent}) and the vanishing of the surface term, we then yield
\begin{eqnarray}
\Delta_{\mu\nu}(p,-p)&=&\int{d^4x}{d^4y}e^{ip\cdot{y}}\theta(x_0)\theta(x_0-y_0)\theta(y_0)\langle\partial_\alpha D_\alpha(x)J_{\mu}(y)J_{\nu}(0)\rangle\nonumber\\
&&+\int{d^4x}{d^4y}e^{ip\cdot{y}}\theta(x_0)\theta(y_0-x_0)\langle J_{\mu}(y)\partial_\alpha D_\alpha(x)J_{\nu}(0)\rangle\nonumber\\
&&+\int{d^4x}{d^4y}e^{ip\cdot{y}}\theta(-x_0)\theta(y_0)\langle J_{\mu}(y)J_{\nu}(0)\partial_\alpha D_\alpha(x)\rangle\nonumber\\
&&+\int{d^4x}{d^4y}e^{ip\cdot{y}}\theta(-x_0)\theta(y_0-x_0)\theta(-y_0)\langle J_{\nu}(0)J_{\mu}(y)\partial_\alpha D_\alpha(x)\rangle\nonumber\\
&&+\int{d^4x}{d^4y}e^{ip\cdot{y}}\theta(-x_0)\theta(x_0-y_0)\langle J_{\nu}(0)\partial_\alpha D_\alpha(x)J_{\mu}(y)\rangle\nonumber\\
&&+\int{d^4x}{d^4y}e^{ip\cdot{y}}\theta(x_0)\theta(-y_0)\langle
\partial_\alpha D_\alpha(x)J_{\nu}(0)J_{\mu}(y)\rangle\nonumber\\
&=&-\int{d^4y}e^{ip\cdot{y}}\theta(y_0)\langle
[D(y_0),J_{\mu}(y)]J_{\nu}(0)\rangle\nonumber\\
&&-\int{d^4y}e^{ip\cdot{y}}\theta(y_0)\langle J_{\mu}(y) [D(0),J_{\nu}(0)]\rangle\nonumber\\
&&-\int{d^4y}e^{ip\cdot{y}}\theta(-y_0)\langle J_{\nu}(0)
[D(y_0),J_{\mu}(y)]\rangle\nonumber\\
&&-\int{d^4y}e^{ip\cdot{y}}\theta(-y_0)\langle
[D(0),J_{\nu}(0)]J_{\mu}(y)\rangle.
\end{eqnarray}
which is simplified, by considering the commutation relation eq.(\ref{charge-generator}), to be
\begin{eqnarray}
\Delta_{\mu\nu}(p,-p)&=&i2d\int{d^4y}e^{ip\cdot{y}}\theta(y_0)\langle T^*(J_{\mu}(y) J_{\nu}(0))\rangle\nonumber\\
& & +i\int{d^4y}e^{ip\cdot{y}}\theta(y_0)\langle
(y\cdot\partial)J_\mu(y)J_{\nu}(0)\rangle+i\int{d^4y}e^{ip\cdot{y}}\theta(-y_0)\langle
J_{\nu}(0) (y\cdot\partial)J_\mu(y)\rangle
\end{eqnarray}
Noticing the following identities that
\begin{eqnarray}
i\int{d^4y}e^{ip\cdot{y}}\theta(y_0)\langle
(y\cdot\partial)J_\mu(y)J_{\nu}(0)\rangle&=&-ip\cdot\frac{\partial}{\partial
p}\int{d^4y}e^{ip\cdot{y}}\theta(y_0)\langle
J_\mu(y)J_{\nu}(0)\rangle\nonumber\\
&&-4i\int{d^4y}e^{ip\cdot{y}}\theta(y_0)\langle
J_\mu(y)J_{\nu}(0)\rangle,
\end{eqnarray}
and
\begin{eqnarray}
i\int{d^4y}e^{ip\cdot{y}}\theta(-y_0)(y\cdot\partial)\langle
J_{\nu}(0) J_\mu(y)\rangle&=&-ip\cdot\frac{\partial}{\partial
p}\int{d^4y}e^{ip\cdot{y}}\theta(-y_0)\langle
J_{\nu}(0) J_\mu(y)\rangle\nonumber\\
&&-4i\int{d^4y}e^{ip\cdot{y}}\theta(-y_0)\langle J_{\nu}(0)
J_\mu(y)\rangle,
\end{eqnarray}
we then obtain the tree-level dilation Ward identity
\begin{eqnarray}
\Delta_{\mu\nu}(p,-p)&=&\{2d-4-p\cdot\frac{\partial}{\partial
p}\}i\int{d^4y}e^{ip\cdot{y}}\langle T^*(
J_\mu(y)J_{\nu}(0))\rangle\nonumber\\
&=&\{2-p\cdot\frac{\partial}{\partial p}\}\Pi^{\mu\nu}(p,-p),
\end{eqnarray}
where $\Pi^{\mu\nu}(p,-p)$ is defined in eq.(\ref{vacuumpolari}) and we have used the classical dimension of $J_\mu$ to be $d=3$.


\end{document}